# Intrinsic Piezoelectric Anisotropy of Tetragonal ABO$_3$ Perovskites: A High-Throughput Study

Fanhao Jia,[1,2] Shaowen Xu,[3] Shunbo Hu,[4,5,1*] Jianguo Chen,[1] Yongchen Wang,[1] Yuan Li,[2] Wei Ren,[1*] and Jinrong Cheng[1*]

[1] *School of Materials Science and Engineering; Department of Physics, International Centre of Quantum and Molecular Structures, Shanghai University, Shanghai 200444, China*
[2] *Department of Physics, School of Sciences, Hangzhou Dianzi University, Hangzhou 310018, China*
[3] *School of Physics and Optoelectronic Engineering, Hangzhou Institute for Advanced Study, University of Chinese Academy of Sciences, Hangzhou 310024, China*
[4] *Institute for the Conservation of Cultural Heritage, School of Cultural Heritage and Information Management, Shanghai University, Shanghai 200444, China*
[5] *Key Laboratory of Silicate Cultural Relics Conservation (Shanghai University), Ministry of Education, China*
*E-mail: shunbohu@shu.edu.cn; renwei@shu.edu.cn; jrcheng@shu.edu.cn

**ABSTRACT**

A comprehensive understand of the intrinsic piezoelectric anisotropy stemming from diverse chemical and physical factors is a key step for the rational design of highly anisotropic materials. We performed high-throughput calculations on tetragonal ABO$_3$ perovskites to investigate the piezoelectricity and the interplay between lattice, displacement, polarization and elasticity. Among the 123 types of perovskites, the structural tetragonality is naturally divided into two categories: normal *tetragonal* (*c/a* ratio < 1.1) and *super-tetragonal* (*c/a* ratio > 1.17), exhibiting distinct ferroelectric, elastic, and piezoelectric properties. Charge analysis revealed the mechanisms underlying polarization saturation and piezoelectricity suppression in the *super-tetragonal* region, which also produces an inherent contradiction between high $d_{33}$ and large piezoelectric anisotropy ratio $|d_{33}/d_{31}|$. The polarization axis and elastic softness direction jointly determine the maximum longitudinal piezoelectric response $d_{33}$ direction. The validity and deficiencies of the widely utilized $|d_{33}/d_{31}|$ ratio for representing piezoelectric anisotropy were reevaluated.

# INTRODUCTION

Over the past century, ABO$_3$-type oxide perovskites encompassing a wide range of compositions are among the most extensively studied functional materials [1], largely owing to their superior ferroelectric [2,3] and piezoelectric [4,5] properties arising from spontaneous polarization [6]. The switchable polarization enables applications in ferroelectric information technologies [7-9], while the piezoelectric conversion of mechanical and electrical energy underlies more broad technologies including ultrasonic transducers and actuators [10,11]. The latter piezoelectricity, as a rank-3 tensor, involves complex coupling between thermal, mechanical, and electrical quantities described by thermodynamics [12].

The tetragonal phase of ABO$_3$ perovskite is of particularly importance among the various structures. Most high piezoelectric response and large ferroelectric polarization in experiments occurred in tetragonal or its mixed phases [13]. Here, we focus on the *P4mm* tetragonal phase (other tetragonal perovskite phases are not ferroelectric or piezoelectric) and a longstanding problem, piezoelectric anisotropy, which is fundamentally important for modern engineering applications like ultrasonic transducers [14] and robotic metamaterials [15]. Highly anisotropic materials can provide minimal noise from lateral vibrations and increased operating frequencies, greatly reducing manufacturing costs through skipping the precise control of thickness resonance mode. Although its importance was acknowledged in the 1990s [16,17], intrinsic factors governing anisotropy such as composition and geometry remain unclear even for those well-studied *P4mm* ABO$_3$. Elucidating underlying mechanisms will facilitate high-anisotropy piezoelectric material design and help the explanation of anomalous properties near the morphotropic phase boundary (MPB) [18] and along nonpolar directions [19,20].

Experimental piezoelectric data are largely lacking for most perovskites. Moreover, reported values vary considerably due to different measurement temperature [21], technique [22] and sample size effects [23]. This motivated us to do a comprehensive study using high-throughput first-principle calculations [24], which is a nascent field endeavoring to speed up time-consuming growth and characterization experiments [25]. In this context, valuable insights into structural distortions, chemical characteristics, polarization, and elastic properties, extent to how they relate to the piezoelectric anisotropy, can be gleaned to obtain an overall picture.

With this aim, we performed density functional perturbation theory (DFPT) calculations on 123 tetragonal $ABO_3$ compounds, including five experimentally synthesized perovskites: $BaTiO_3$ (BTO), $KNbO_3$ (KNO), $PbTiO_3$ (PTO), $BiCoO_3$ (BCO) and $PbVO_3$ (PVO). Other perovskites including $KTaO_3$ [26] and $SnTiO_3$ [27], which can potentially be synthesized as strained thin films or superlattices, were not considered as already synthesized crystals in this work. Studies on lattice deformation and interatomic displacement (Figure 1a) showed that the latter dominates structural tetragonality. The tetragonal phase stabilization energy, displacement amplitude, polarization, elastic, and piezoelectric properties naturally cluster into normal *tetragonal* (lattice anisotropy *c/a* ratio < 1.1) and *super-tetragonal* (*c/a* ratio > 1.17) groups with distinct chemical features. Polarization analysis revealed that its saturation in *super-tetragonal* perovskites is related to the charge transfer limit between atomic layer, which also results in the strong suppression of piezoelectric response when *c/a* ratio >1.3.

The piezoelectric response is dependent on the selected crystallographic coordinate system, while longitudinal surfaces provide a classical way to visualize the piezoelectric anisotropy [28]. For tetragonal perovskites, we found two classes of longitudinal piezoelectric response shapes, which are closely related to the polarization and elastic softness. Due to the constitutive law, the elastic anisotropy significantly impacts piezoelectric anisotropy. The direction of maximum longitudinal response $d_{33}$ is determined by the softest elasticity direction and polarization axis in conjunction. Among the studied perovskites, PTO exhibits the largest $d_{33}$ from several favorable conditions, including the moderate polarization switching energy, the softest elastic properties among *tetragonal* perovskites, coincidence of polarization, softest elasticity, and maximum $d_{33}$ directions. Our high-throughput calculations reveal the challenge in obtaining both large $d_{33}$ and high $|d_{33}/d_{31}|$ anisotropy concurrently. The dimensionless parameter $|d_{33}/d_{31}|$ ratio is widely used in the literature to characterize piezoelectric anisotropy [29-31], which in fact lacks rigor for most perovskites, as $d_{31}$ is a quantity that in principle depends both on the longitudinal polarization and lateral stresses. $|d_{33}/d_{31}|$ ratio is a constant only works for 29 perovskites along the particular polarization direction, whose polarization and maximum $d_{33}$ directions are identical.

The following section firstly provides an overview of the high-throughput calculation framework and the quantities to be considered in this work. The discussion begins with an analysis of lattice anisotropy and spontaneous symmetry breaking to introduce two groups of *P4mm* perovskites, namely *tetragonal* and *super-tetragonal*. Subsequently, we connect polarization change to interatomic displacement and Born Effective Charges (BECs) through detailed analysis. We then address elastic anisotropy and visualize the three-dimensional (3D) Young's modulus to portray the varying elastic softness. In the final section, we introduce the conditions for large piezoelectric coefficient $d_{33}$, analyze the reasons for suppressed response in *super-tetragonal* perovskites, and revisit the physical meaning and validity of piezoelectric anisotropy parameter $|d_{33}/d_{31}|$ and its limitations for tetragonal perovskites. The Methods section provides additional computational details not covered in preceding sections.

**RESULTS AND DISCUSSION**

**High-throughput calculation and screening**

Our high-throughput calculation and data screening workflow are illustrated in Figure 1b. The Material Project database contains over 15,000 experimental synthesized or theoretically predicted crystals [32], from which we conducted initial screening by excluding certain unlike elements and selecting compounds with low energy above the hull. Among the screened 988 ABO$_3$ compounds, 825 matched the symmetry conditions of the perovskite structure. Each compound may have multiple structural phases, depending on ferroelectric displacements and oxygen octahedral rotations, which can be metallic or insulating as well. We selected 284 ABO$_3$ perovskites possessing at least one insulator phase with band gap $E_g > 0.1$ eV for further consideration, among which 145 are synthesized and 20 are ferroelectric in experiments. Notably, our screening successfully identified all five experimentally crystallized *P4mm* perovskites: the nonmagnetic BTO, KNO, and PTO as well as the antiferromagnetic BCO and PVO.

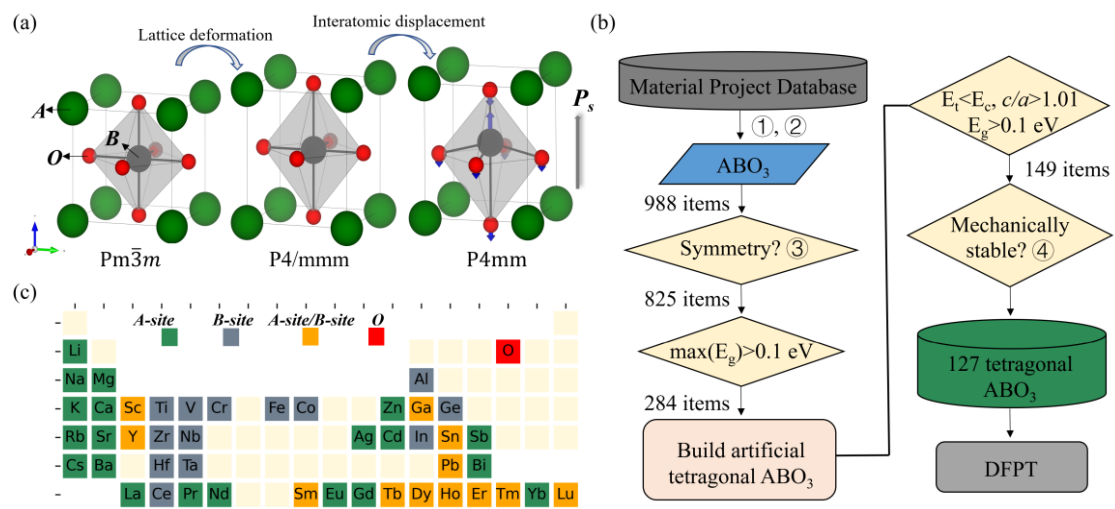

Figure 1. **Crystal lattice and high-throughput workflow.** (a) Cubic and tetragonal ABO₃ perovskite structures, showing lattice deformation and interatomic displacement. (b) Workflow of the high-throughput calculation and data screening by selection rules: ① excluding H, B, Si, C, N, P, I, Br, F, Cl, S, Se, Te; ② energy above hull ≤ 1 eV/atom; ③ space group matches perovskites: No. 11, 12, 14, 15, 26, 33, 38, 62, 63, 74, 99, 123, 127, 140, 160, 161, 167, 204, 221; Wyckoff positions match with prototype perovskites; ④ $c_{11} > |c_{12}|, 2c_{13}^2 < c_{33}(c_{11} + c_{12}), c_{44} > 0, c_{66} > 0$; Band gap $E_g > 0.1$ eV and relative energy between tetragonal ($E_t$) and cubic ($E_c$) phases. (c) Distributions of A- and B-sites elements for 127 selected tetragonal ABO₃ systems.

The high-throughput calculations were performed on the screened set of 284 ABO₃ perovskites. We artificially constructed their cubic and tetragonal structures to calculate relative stabilities and electronic properties. We selected 149 ABO₃ systems for which the tetragonal phase exhibited lower total energy ($E_t < E_c$), *c/a* ratio >1.01, and $E_g > 0.1$ eV. The mechanical stabilities of these artificial *P4mm* structures were then evaluated, which is important to ensure the viability of elasticity that was used to obtain piezoelectric tensor [28,33]. Finally, 127 tetragonal *P4mm* ABO₃ perovskites were screened, with the distributions of A-site and B-site elements shown in Figure 1c. As expected, alkali metals predominantly occupied A-sites while transition metals were common on B-sites; rare earth elements could occupy both sites. DFPT calculations were conducted on these 127 tetragonal *P4mm* ABO$_3$ perovskites to obtain their ferroelectric, mechanical, and piezoelectric properties.

**Lattice tetragonality and interatomic displacement**

Figure 1a illustrates the perovskite structures, including the ideal cubic phase (space group *Pm$\bar{3}$m*) and two tetragonal phases (*P4/mmm* and *P4mm*). The paraelectric cubic phase represents the high-temperature phase of most perovskites, distorting upon cooling to room temperature. The *P4/mmm* phase involves only a lattice deformation with lattice constants $a = b < c$, known as the lattice tetragonality. The $c/a$ ratio is widely used to quantify this tetragonality. The *P4mm* phase includes both lattice deformation and additional uniaxial interatomic displacements along the *c* axis, breaking structural inversion symmetry and enabling key functionalities like ferroelectricity and piezoelectricity. Table I lists the structural parameters for five experimentally synthesized *P4mm* perovskites. The interatomic displacement $\Delta r_\alpha$ ($\alpha$ is ion index) occurs under a symmetry mode of $\Gamma_3^-$, whose amplitude can be represented by $u = \sqrt{\sum_\alpha \Delta r_\alpha^2}$.

Both lattice deformation and interatomic displacement are symmetry breaking caused by lattice instability stemming from Jahn-Teller or pseudo-Jahn-Teller effects [34]. However, whether these two major structural instabilities in tetragonal perovskites are both spontaneous was usually well undistinguished. Analysis of the 127 ABO$_3$ perovskites herein reveals lattice deformation to be mostly nonspontaneous. Only four *P4/mmm* ACrO$_3$ possess lower total energy than the cubic phase, while most *P4/mmm* ABO$_3$ in our calculation revert to cubic during the lattice optimization. As shown in Figure 2a, when *P4/mmm*-phase lattices were fixed to the P4mm lattice constants, the total energy increases significantly, especially those with $c/a$ ratio > 1.17. This further signifies lattice deformation as generally unfavorable for structural tetragonality. In contrast, interatomic displacement is spontaneous for most tetragonal perovskites, except the four ACrO₃. Figure 2b shows the tetragonal stabilization energy ($E_t$-$E_c$) to be proportional to displacement amplitude, indicating that distortions in most tetragonal ABO₃ are indeed dominated by interatomic displacements (or optical phonons), rather than the lattice deformation. The latter mostly occurs passively due to the bond length changes caused by former, as the positive relationship shown in Figure 2c.

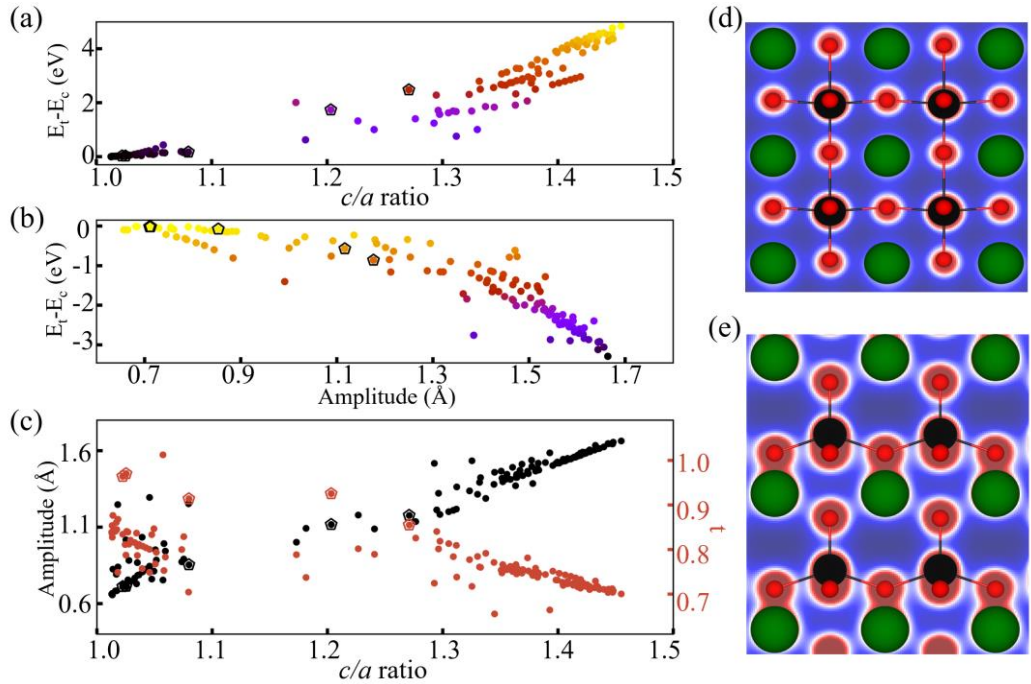

Figure 2. **Structural instabilities and distribution.** (a) Energy differences between the tetragonal *P4/mmm* phase (with lattice constants fixed to the *P4mm* phase) and the cubic phase ($E_t$-$E_c$) as a function of *c/a* ratio. (b) Energy differences between the tetragonal *P4mm* and cubic phases ($E_t$-$E_c$) versus interatomic displacement amplitude. (c) Distributions of displacement amplitude and Goldschmidt tolerance factor with respect to the *c/a* ratio. Pentagons highlight five experimentally synthesized *P4mm* perovskites. Planar charge densities of (d) *tetragonal* BTO and (e) *super-tetragonal* BCO.

The two structural instabilities also differ in their distributions, as shown in Figure 2a and 2b. The interatomic displacement, represented by its amplitude, is continuously distributed over a range from 0.66 to 1.66 Å. In contrast, the *c/a* ratio distribution and its impacts on total energy and amplitude naturally divide into two categories. The normal *tetragonal* perovskites like BTO, KNO and PTO belong to the first category with *c/a* ratio < 1.1, while PVO and BCO comprise the second *super-tetragonal* category with *c/a* ratio >1.17. It's worth mentioning that BCO and PVO were synthesized under high pressure [35,36], indicating more difficult preparation compared to conventional perovskites. From the perspective of ferroelectricity, the latter may be also known as *hyperferroelectrics* [37] or pyroelectric materials [36] due to their persistent polarizations and extremely high ferroelectric switching energies. The discontinuity of *c/a* ratio distribution does not result solely from the ionic radius difference between A-site and B-site ions (see the Goldschmidt tolerance factor distribution in Figure 2c). Although a smaller tolerance factor does promote greater lattice tetragonality in the *super-tetragonal* region, it can't fully account for the observed discontinuity. Beside the ionic reason, the discontinuous super tetragonality more likely results from the chemical side change, which involves a combination of one B-O bond breaking and A-O bond strengthening, as depicted in Figure 2d and 2e.

Table I. Structural properties of five experimentally synthesized tetragonal $ABO_3$ perovskites, including the lattice constant $a$ (Å), $c/a$ ratio, and interatomic displacement amplitude $u$ (Å). Also listed are the spontaneous polarization $P$ ($\mu C/cm^2$) and piezoelectric constants $d_{33}$ and $d_{31}$ ($pC/N$) and their ratio $|d_{33}/d_{31}|$.

| System | $a$ | $c/a$ | $u$ | $P$ | $d_{33}$ | $d_{31}$ | $|d_{33}/d_{31}|$ |
|---|---|---|---|---|---|---|---|
| BaTiO$_3$ | 3.968 | 1.025 | 0.712 | 30 | 49.0 | -9.2 | 5.3 |
|  | 3.995[a] | 1.010[a] |  | 26[a1] | 69[a2] | -33[a2] | 2.1[a2] |
| KNbO$_3$ | 3.970 | 1.022 | 0.710 | 31 | 40.1 | -7.7 | 5.2 |
|  | 3.997[b] | 1.017[b] |  | 37[b1] |  |  |  |
| PbTiO$_3$ | 3.878 | 1.078 | 0.850 | 85 | 209.0 | -40.3 | 5.2 |
|  | 3.895[c] | 1.071[c] |  | 75[c1] | 137[c2] | -22.8[c2] | 6.0[c2] |
| PbVO$_3$ | 3.810 | 1.203 | 1.120 | 129 | 120.9 | -17.1 | 7.1 |
|  | 3.800[d] | 1.229[d] |  |  |  |  |  |
| BiCoO$_3$ | 3.662 | 1.271 | 1.177 | 160 | 26.7 | 8.7 | 3.1 |
|  | 3.729[e] | 1.266[e] |  |  |  |  |  |

[a]Ref [38], [a1]Ref [39], [a2]Ref [40]; [b]Ref [41], [b1]Ref [42]; [c]Ref[43], [c1]Ref [44], [c2]Ref [22]; [d]Ref [36]; [e]Ref [35]

**Spontaneous polarization**

The spontaneous polarization of tetragonal perovskites lies in the substantial interatomic displacements, which alter the distribution of valence charges through relaxations of ion positions and the rehybridization of the atomic orbitals. The macroscopic polarization value can be obtained by summing the products of the BECs $Z_{\alpha m}$ and the ionic relative displacement $\boldsymbol{u}_{\alpha m}$: $P = 1/\Omega \sum_{\alpha m} \boldsymbol{Z}^*_{\alpha m} \Delta \boldsymbol{u}_{\alpha m}$, where $\Omega$ is cell volume. The calculated polarization values listed in Table I agrees with those derived via Berry-phase method [45,46]. Four ACrO3 perovskites were abandoned in the subsequent discussion because they had no interatomic displacement, *i.e.,* we are now left with only 123 tetragonal ABO3 perovskites.

As expected, Figure 3a and 3b have illustrated that polarization generally increases with larger displacement amplitude and *c/a* ratio. However, when *c/a* ratio > 1.3, the polarization increase tends to saturate, despite continued growth in amplitude. This convergence cannot be attributed to cancelation effects among the A-site, B-site, and O ions. As shown in Figure 3c, all polarization contributions from different ions increase overall with the *c/a* ratio and begin to converge when the *c/a* ratio surpasses 1.3. The polarization saturation stems from the BEC side. Indeed, as shown in Figure S1, the BECs trend toward convergence as the amplitude rises for *c/a* ratio > 1.3. With a converged BEC, the polarization increase from a larger amplitude is counteracted by the concomitant increase in volume, since polarization is inversely related to volume (Figure S2). In other words, as the *c/a* ratio grows, charge transfer between atomic layers becomes hindered, and the system approaches periodic parallel plate capacitor with a fixed charge amount on each plate. When the planar charge is converged, the polarization of the capacitor no longer climbs as the plate separation expands.

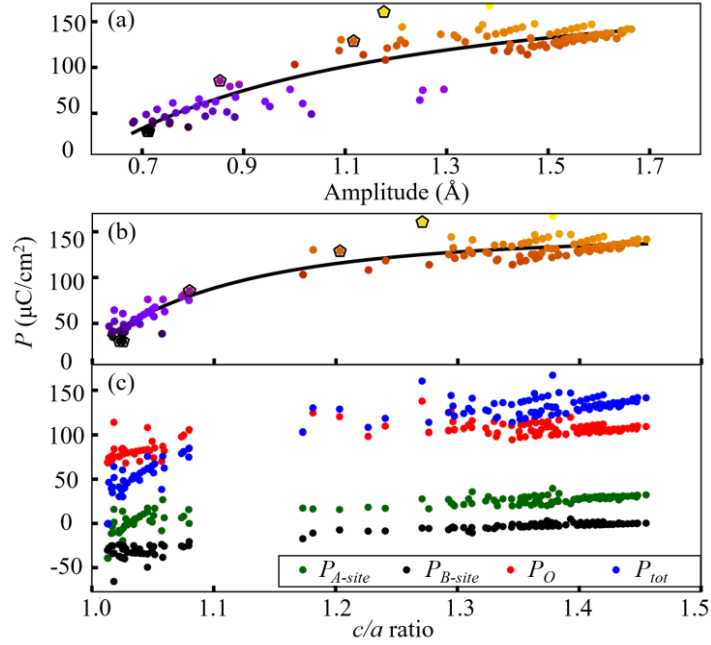

Figure 3. **Spontaneous polarization analysis.** (a) Polarization ($P$) versus interatomic displacement amplitude. (b) Polarization and (c) the contributions from A-site, B-site, and O ions as a function of $c/a$ ratio.

**Elastic anisotropy**

Due to the constitutive law, elastic anisotropy is intimately connected to piezoelectric anisotropy. The presence of polarization and lattice tetragonality renders the elasticity along the $c$-axis more compliant than the $a$- or $b$-axes in tetragonal perovskites, as evidenced by longitudinal stiffness anisotropy ratios $c_{33}/c_{11}$ less than 1.0, where $c_{33}$ and $c_{11}$ are two independent longitudinal elements of the elastic stiffness matrix. Figure 4a and 4b compare $c_{33}$ and $c_{11}$ as a function of $c/a$ ratio. $c_{11}$ generally decreases with increasing $c/a$ ratio, while $c_{33}$ initially decreases for $c/a$ ratio below 1.1 but transitions to an increase when $c/a$ ratio > 1.17. This suggests a qualitative change in the $c$-axis stiffness between *tetragonal* and *super-tetragonal* perovskites, which directly impacts piezoelectric anisotropy.

Figure 4c displays the $c_{33}/c_{11}$ as a function of $c/a$ ratio. In the *tetragonal* region, there are perovskites with $c_{33}/c_{11}$ approaching both 1.0 and 0.1. For perovskites like rare earth chromium acids independent to the $c/a$ ratio present small longitudinal stiffness anisotropy, namely their $c_{33}/c_{11}$ ratios approach 1.0. While the longitudinal stiffness anisotropies of others are proportional to the $c/a$ ratio. For perovskites like

PTO, the softness of A-site elements makes $c_{33}$ much smaller than $c_{11}$, thus presenting much larger stiffness anisotropy in *tetragonal* perovskites. In the *super-tetragonal* regions, due to the opposite changing trends of $c_{33}$ and $c_{11}$, the longitudinal stiffness anisotropy gradually decreases as the *c/a* ratio increases. When *c/a* ratio >1.4, perovskites approach isotropy, suggesting an anomalous inverse relationship between lattice anisotropy and the elastic anisotropy for these *super-tetragonal* perovskites.

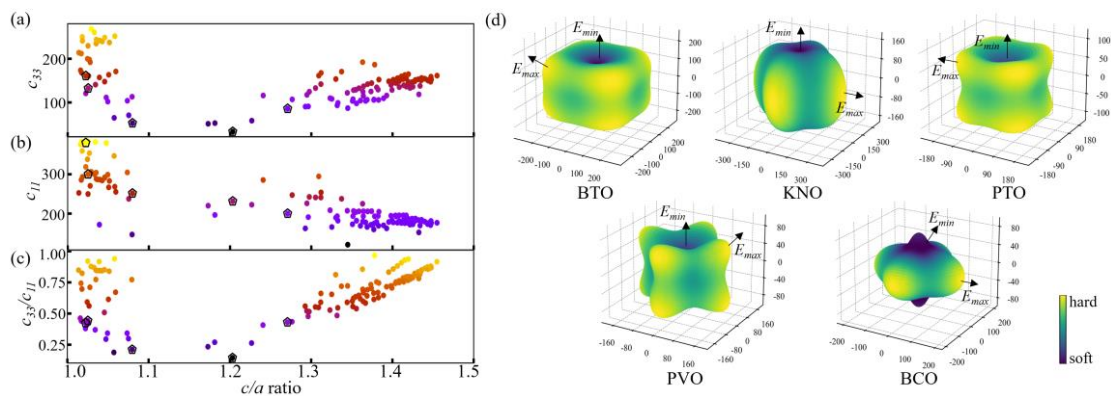

Figure 4. **Elastic anisotropy.** (a-c) elastic constants as a function of *c/a* ratio. (d) 3D representations of Young's moduli of five experimental *P4mm* perovskites.

Another more intuitive representation of elastic anisotropy employs 3D renderings of the directionally dependent elastic moduli, rigorously defined by the elastic compliance tensor $S$ (the inverse of the stiffness tensor). Due to the orientation of chemical bonds, all elastic moduli of single crystal are unavoidably directionally dependent. The values along the lattice vectors for five experimentally synthesized P4mm perovskites are listed in Table SII. Figure 4d shows their 3D Young's moduli ($E$), which measures the degree of a crystal extends or shortens under tension/compression. Among them, BTO, KNO, PTO, and PVO exhibit minimum $E$ along the *c* axis, aligned with the polarization axis. In contrast, BCO has a minimum $E$ tilted 35° from the normal polarization axis. For most *super-tetragonal* perovskites, they adopt the BCO-like elastic softness. These two different elastic softness characteristics are intrinsic. It is worth mentioning that other extrinsic factors, such as temperature and strain near the phase boundaries [21], can lead to elastic softness direction change as well.

**Piezoelectric anisotropy**

Piezoelectricity refers to the response of polarization on the external strain or stress or, equivalently, the strain/stress response on an external electric field. The tensor $d$ is sometimes termed a "piezoelectric strain constant", more commonly used in the experimental literature than "piezoelectric stress constant" $e$. The latter is more conveniently obtained from the first-principle calculations, which is convertible via the relation: $e_{\alpha j} = C_{jk} d_{\alpha k}$. This rank-3 tensor contains three independent constants according to *P4mm* symmetry (in Voigt notation):

$$d = \begin{bmatrix} 0 & 0 & 0 & 0 & d_{15} & 0 \\ 0 & 0 & 0 & d_{15} & 0 & 0 \\ d_{31} & d_{31} & d_{33} & 0 & 0 & 0 \end{bmatrix}. \quad (1)$$

$d_{33}$ denotes the change of polarization $P_3$ due to a longitudinal stress $\sigma_3$, while $d_{31}$ refers to the change in $P_3$ from a lateral stress $\sigma_1$. Table I provides the specific $d_{33}$ and $d_{31}$ values of five experimentally synthesized *P4mm* perovskites. Figure 5a shows $d_{33}$ and $d_{31}$ as a function of *c/a* ratio. Among the 123 *P4mm* ferroelectric perovskites, PTO exhibits the largest $d_{33}$ of 209 *p*C/N, attributed to: (i) it's located in the *tetragonal* region with a moderate ferroelectric switching energy ($E_{P4/mmm}$-$E_{P4mm}$ < 0.3 eV) but substantial polarization (Figure S3), indicating facile polarization rotation in an electric field; (ii) the softest elastic properties along the polarization axis among the *tetragonal* perovskites; and (iii) alignment of its minimum $E$ and maximum $d_{33}$ directions with the polarization axis, enabling optimal response deliver. Systems in the *super-tetragonal* region with *c/a* ratio ≤ 1.2 can also present large $d_{33}$ (>100 *p*C/N), because they partially satisfy the criteria for very large $d_{33}$. For instance, PVO has a much larger ferroelectric switching energy (>1 eV), but it has the softest elastic properties among all 123 perovskites, as well as the aligned minimum $E$ and maximum $d_{33}$ directions along the polarization axis. On the other hand, when *c/a* ratio >1.3, $d_{33}$ and $d_{31}$ will approach very small due to the polarization passivation from charge transfer limitation and their harder elastic properties, as we discussed in the previous sections. Similar patterns also happen for the constants $e_{33}$ and $e_{31}$ as shown in Figure S4.

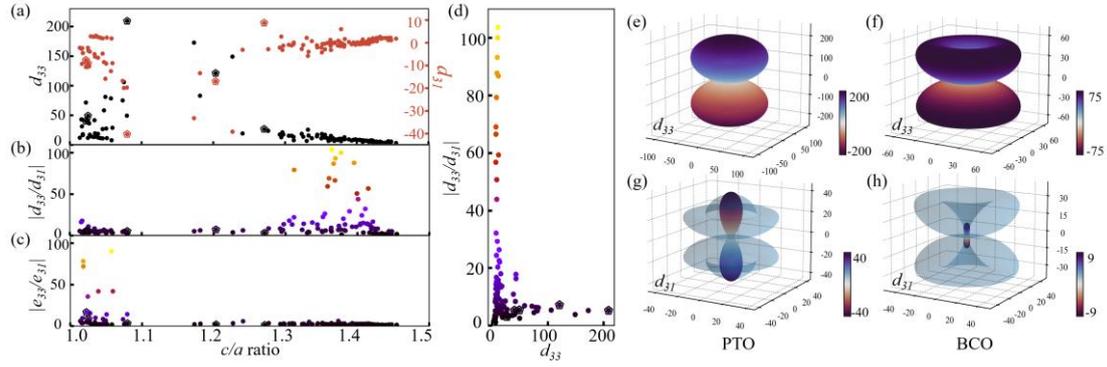

Figure 5. **Piezoelectricity anisotropy**. (a) Piezoelectric constants $d_{33}$ and $d_{31}$, (b) $|d_{33}/d_{31}|$ and (c) $|e_{33}/e_{31}|$ ratios as a function of c/a ratio. The 3D representations of piezoelectricity of *P4mm* perovskites: (e-f) surfaces of $d_{33}$, namely under longitudinal stress. (g-h) surfaces of $d_{31}$ within Marmier manner [47], longitudinal polarization under varying lateral stress.

In literature, the ratio $|d_{33}/d_{31}|$ or $|e_{33}/e_{31}|$ were widely used to represent piezoelectric anisotropy. Figure 5b and 5c display these two anisotropy ratios as a function of *c/a* ratio. These are some major differences between the ***d*** and ***e***. In the *tetragonal* region, most perovskites present a small $|d_{33}/d_{31}|$ ratio <10. In the *super-tetragonal* region, especially for *c/a* ratio >1.3, some perovskites display very large $|d_{33}/d_{31}|$ ratio due to their $d_{31}$ approach zero. On the contrary, perovskites with large $|e_{33}/e_{31}|$ ratio are in the *tetragonal* region due to their $e_{31}$ approach zero, as illustrated in Figure S4. In the *super-tetragonal* region, both $e_{33}$ and $e_{31}$ will slowly approach 0.5 C/m$^2$, because when the BEC converge, longitudinal and lateral strains have the same effect on the volume change, the polarization changes are close and small. Correspondingly, we could understand why *P4mm* perovskites can achieve both high $|e_{33}/e_{31}|$ anisotropy and large $e_{33}$ response (Figure S5), but cannot realize a high $|d_{33}/d_{31}|$ anisotropy simultaneously with a large $d_{33}$ response (Figure 5d). In the following, we primarily focus on the more commonly used piezoelectric constant ***d***.

Since both $d_{33}$ and $d_{31}$ are directionally dependent, the most classical method to visualize piezoelectric anisotropy is the longitudinal surface [28]. The longitudinal surface of infinitesimal area is normal to an axial tensile stress ***σ′*** parallel to ***a′***. The polarization ***P′*** normal to the surface is given by ***P′*** = ***d′σ′***, where ***d′*** is the polarization value (response) parallel to *a'* for unit stress. The longitudinal surface value *d(a')* is specified by the radius vector *a'*, previously oriented along *a*. For example,

$$d'_{333} = a_{3i}a_{3j}a_{3k}d_{ijk}. \tag{2}$$

In addition, the longitudinal polarization along *a* can also response to lateral stress *b* as well. For example,

$$d'_{311} = a_{3i}b_{1j}b_{1k}d_{ijk}, \tag{3}$$

where $a_{ij}$ and $b_{ij}$ are both direction cosines.

Utilizing these equations, we can calculate the 3D representation surfaces for $d_{33}$ and $d_{31}$. The response value may be positive or negative, depending on whether the polarization change is parallel or anti-parallel to the polarization direction. There are two substantially different types of longitudinal surface of $d_{33}$, as displayed in Figure 5e and 5f for PTO and BCO, respectively. PTO's $d_{33}$ resembles two adhered spheres, while BCO's appears as two attached toruses. In PTO, the maximum $d_{33}$ aligns with the *c* axis, coinciding with the polarization and minimum *E* directions. The minimum $d_{33}$ lies in the *ab* plane. 29 perovskites exhibit a similar $d_{33}$ shape to PTO, which are all in a region with *c/a* ratio < 1.4. Most perovskites resemble BCO-like $d_{33}$ shape. Their minimum $d_{33}$ is also in the *ab* plane. However, the maximum $d_{33}$ direction no longer coincides with the polarization axis, tilted 51° away, even further than minimum *E* direction. A common feature among perovskites with off-axis $d_{33}$ maximum is that they usually have a deviation of their minimum *E* from the polarization axis. In BCO, the $d_{33}$ along the polarization direction is just 26.7 *p*C/N, much less than the 74.6 *p*C/N maximum.

The $d_{31}$ shapes are similar across all perovskites, as displayed in Figure 5g and 5h in the Marmier manner [47]. The inner solid surface denotes the minimum $d_{31}$ along this longitudinal direction, which can exhibit negative piezoelectricity as in PTO. The outer transparent surface represents the maximum $d_{31}$ along the same longitudinal direction. The overall maximum $d_{31}$ direction (transparent surface) always deviates from the polarization axis. On the other hand, only when the longitudinal direction is set to the polarization direction, the maximum $d_{31}$ and minimum $d_{31}$ are constants, which are identical in this direction.

In the last, let's revisit more rigorously the $|d_{33}/d_{31}|$ ratio widely utilized in literature to characterize the piezoelectric anisotropy of tetragonal perovskites. For PTO-like single-crystal perovskites (Figure 5e), this definition makes sense when along the polarization direction. Since the maximum $d_{33}$ aligns with the polarization axis, it's a constant. Moreover, the $d_{31}$ is also a constant (maximum and minimum $d_{31}$ equal) when the applied lateral stress is perpendicular to the polarization direction, independent on the lateral orientation. Hence, $|d_{33}/d_{31}|$ constitutes a suitable constant quantifying the ratio between the overall maximum response from longitudinal stress and the response along this longitudinal axis but from lateral stress. This also applies for polarized tetragonal polycrystalline and ceramic systems with maximum $d_{33}$ aligning with the overall polarization direction. They are isotropic in the plane perpendicular to the polarization, so their $d_{31}$ is likewise constant along the polarization.

On the contrary, for BCO-like single-crystal perovskites, the maximum $d_{33}$ lies on a ring tilted from the polarization axis, as depicted in Figure 5f. Meanwhile, $d_{31}$ is only constant when the longitudinal stress is polarization-aligned. Although $|d_{33}/d_{31}|$ along the polarization axis is still a constant, but it doesn't represent the direction of overall maximum $d_{33}$. Moreover, for other longitudinal orientations, $|d_{33}/d_{31}|$ is a function depends on the lateral stress. Therefore, for BCO-like tetragonal single-crystal perovskites, $|d_{33}/d_{31}|$ is no longer an adequate parameter for characterizing piezoelectric anisotropy. The same issue should also exist for polycrystalline and ceramic systems lacking alignment between overall polarization axis and directions of maximum piezoelectric response.

**CONCLUSION**

High-throughput calculations were conducted for *P4mm* ABO$_3$ perovskites to comprehensively elucidate the mechanisms governing intrinsic piezoelectric anisotropy spanning diverse chemical and physical factors. Interatomic displacement drives the spontaneous structural distortion. Lattice anisotropy divides the 123 ABO$_3$ perovskites into two groups: *tetragonal* and *super-tetragonal*, exhibiting distinct characteristics in ferroelectric switching barrier, polarization, elasticity, and piezoelectricity. Polarization saturation and piezoelectric response suppression in the *super-tetragonal* region result from the charge transfer limit, which also produces an inherent contradiction between high $d_{33}$ and large piezoelectric anisotropy ratio $|d_{33}/d_{31}|$. The elastic softness direction and polarization axis jointly control the 3D longitudinal piezoelectric response, with the former playing the major role in determining the orientation of maximum longitudinal piezoelectric response. The piezoelectric coefficient ratio $|d_{33}/d_{31}|$ cannot rigorously represent the anisotropy of most tetragonal single-crystal perovskites. The condition for $|d_{33}/d_{31}|$ utilization as a constant anisotropy calibrator along the polarization direction, such as in BTO/PTO-like systems, is alignment between the polarization axis and maximum longitudinal piezoelectric response. Overall, we hope this study serves as a useful foundation toward rational design of intrinsically highly anisotropic piezoelectric materials. These important intrinsic factors should also be considered in multiscale modeling of all piezoelectric systems.

**METHODS**

First-principles calculations were carried out using in the Vienna *ab initio* Simulation Package (VASP) [48,49] within the density functional theory (DFT) framework. The generalized gradient approximation (GGA) exchange-correlation functional in a version of the Perdew-Burke-Ernzerhof formulation optimized for solids (PBEsol) was used [50]. An 8×8×6 Γ-centered *k* grid [51], a plane-wave energy cutoff of 550 eV, and the Gaussian smearing method with a width of 0.05 eV were adopted.

The elastic constants, piezoelectric constants, and Born effective charge determined via the DFPT routines [52], briefly summarized as follows [53]. When perturbations (displacements $u_m$, strains $\eta_j$ or electric field $\varepsilon$) are applied to a crystal, the first-order responses to the total energy $E$ are (i) forces $F_m$, (ii) stresses $\sigma_j$, and (iii) polarizations $P_\alpha$. The pertinent second derivatives of the total energy per unit volume can then be constructed: elastic stiffness constants $C_{ij}=d\sigma_i/d\eta_j$, Born effective charges $Z_{\alpha m}=dP_\alpha/du_m$, and piezoelectric stiffness constants $e_{\alpha j}=dP_\alpha/d\eta_j$. The $e_{\alpha j}$ describes the change of polarization $P_\alpha$ arising from a strain $\eta_j$, which equals to the stress $\sigma_j$ from a change of homogeneous electric field $\varepsilon_\alpha$. Another equivalent definition, more commonly used in experimental literatures, is the $d_{\alpha j}=d\eta_j/d\varepsilon_\alpha$ that describes the change of strain $\eta_j$ from a field $\varepsilon_\alpha$, or polarization $P_\alpha$ change arising from a stress $\sigma_j$. $e_{\alpha j}$ and $d_{\alpha j}$ differ physically, they can be easily converted using elastic constants: $e_{\alpha j}= C_{ij}d_{\alpha j}$ vs. $d_{\alpha j}= S_{ij}e_{\alpha j}$. Here, $S_{ij}=d\eta_j/d\sigma_j$ are the elastic compliance constants, related by $S_{ij}=C_{ij}^{-1}$. Computationally, $C_{ij}$ and $e_{\alpha j}$ are more straightforward to obtain than the $S_{ij}$ and $d_{\alpha j}$. The visualizations of elasticity and piezoelectricity were based on the framework of ELATE [54].

## ACKNOWLEDGEMENTS

This work was supported by China Postdoctoral Science Foundation (No. 2022M722035); National Natural Science Foundation of China (No. 11929401, 52271007, 12274278, 12074241, and 52130204); the Original exploration project of Shanghai Natural Science Foundation (No. 22ZR1481100), the Young Scientists Fund of National Natural Science Foundation of China (No. 12204300); Science and Technology Commission of Shanghai Municipality (No. 22XD1400900, 20501130600, 21JC1402600, and 21JC1402700); the Key Research Project of Zhejiang Lab (No. 2021PE0AC02); the Major Science and Technology Projects of Shanxi Province (No. 202201150501024); High Performance Computing Center, Shanghai Technical Service Center of Science and Engineering Computing, Shanghai University. Dr. Jia also appreciated the discussion with P.Z.

## AUTHOR CONTRIBUTIONS

F.J. carried out the DFT calculations under the supervision of W.R and J.C. All authors contributed to the discussion and writing of manuscript. W.R and J.C. conceived and led the entire scientific project.

## COMPETING INTERESTS

The authors declare no competing interests.